# Modeling the impact of spatial oxygen heterogeneity on radiolytic oxygen depletion during FLASH radiotherapy

Edward Taylor[1,2], Richard P. Hill [†,1,2,3], Daniel Létourneau[1,2].

[1]Radiation Medicine Program, Princess Margaret Cancer Centre, Toronto, Ontario, Canada
[2]Department of Radiation Oncology, University of Toronto, Toronto, Ontario, Canada
[3]Department of Medical Biophysics, University of Toronto, Toronto, Ontario, Canada

## Abstract

**Purpose.** It has been postulated that the delivery of radiotherapy at ultra-high dose rates ("FLASH") reduces normal tissue toxicities by depleting them of oxygen. The fraction of normal tissue and cancer cells surviving radiotherapy depends on dose and oxygen levels in an exponential manner and even a very small fraction of tissue at low oxygen levels can determine radiotherapy response. To quantify the differential impact of FLASH radiotherapy on normal and tumour tissues, the spatial heterogeneity of oxygenation in tissue should thus be accounted for.

**Methods.** The effect of FLASH on radiation-induced normal and tumour tissue cell killing was studied by simulating oxygen diffusion, metabolism, and radiolytic oxygen depletion over domains with simulated capillary architectures. To study the impact of heterogeneity, two architectural models were used: 1.) randomly distributed capillaries and 2.) capillaries forming a regular square lattice array. The resulting oxygen partial pressure distribution histograms were used to simulate normal and tumour tissue cell survival using the linear quadratic model of cell survival, modified to incorporate oxygen-enhancement ratio (OER) effects. The ratio ("dose modifying factors") of conventional low-dose-rate dose and FLASH dose at iso-cell survival was computed and compared with empirical iso-toxicity dose ratios.

**Results.** Tumour cell survival was found to be increased by FLASH as compared to conventional radiotherapy, with a 0-1 order of magnitude increase for expected levels of tumour hypoxia, depending on the relative magnitudes of radiolytic oxygen depletion and tissue oxygen metabolism. Interestingly, for the random capillary model, the impact of FLASH on well-oxygenated (normal) tissues was found to be much greater, with an estimated increase in cell survival by up to 10 orders of magnitude, even though reductions in mean tissue partial pressure were modest, less than ~7 mmHg for the parameter values studied. The dose modifying factor for normal tissues was found to lie in the range 1.2-1.7 for a representative value of normal tissue oxygen metabolic rate, consistent with preclinical iso-toxicity results.

**Conclusions.** The presence of very small nearly hypoxic regions in otherwise well-perfused normal tissues with high mean oxygen levels resulted in a greater proportional sparing of normal tissue than tumour cells during FLASH irradiation, possibly explaining empirical normal tissue sparing and iso-tumour control results.

## 1. Introduction

Within the oxygen depletion hypothesis, it is commonly argued that FLASH radiotherapy reduces normal tissue toxicity by decreasing the level of oxygen in these tissues, thus rendering them more radioresistant due to the OER effect [1-5]. Low linear-energy-transfer (LET) radiation damages DNA primarily through the generation of free radicals. Oxygen is needed to "fix" this damage, rendering it irreparable; the lower the oxygen levels in tissue, the lower the impact of radiation [6, 7]. This effect is usually quantified by the OER, defined as the ratio $D_{\text{hypoxic}}/D(p)$, where $D_{\text{hypoxic}}$ is the dose needed to achieve a given level of cell killing under maximally hypoxic conditions and $D(p)$ is the dose needed to achieve the same level of cell killing at an oxygen partial pressure $p$. Viewed as a function of partial pressure, the OER is equal to one at zero tissue partial pressure, $p = 0$. As the oxygen partial pressure is increased, OER($p$) initially increases very rapidly, nearly reaching its maximum value (between 2.5 and 3) by 10-15 mmHg [8, 9], but changes slowly above this range. The mean oxygen partial pressures $\bar{p}$ of most normal tissues range from ~20 to 50 mmHg [10]. Based on the slow evolution of OER($p$) with respect to $p$ at partial pressures above ~ 10 mmHg, one might thus expect that a substantial ($\gtrsim$ 10-40 mmHg) reduction in partial pressure by radiation would be needed to significantly alter cell killing in normal tissues and hence, reduce normal tissue toxicities. Recently, this has been used to argue that the oxygen depleting effects of FLASH cannot be responsible for its apparent reductions in toxicity, due to comparatively small reductions in oxygen levels [11, 12].

---

[†] Dr. Hill died during preparation of this manuscript.

From both *in vitro* [13-15] and *in vivo* [16] experiments, radiation has been shown to deplete oxygen in a dose-dependent manner by 0.1–0.42 mmHg per Gy of delivered radiation, defining the rate

$$G \equiv \Delta\bar{p}/D \qquad (1)$$

of radiolytic oxygen depletion (ROD). Here, $\Delta\bar{p}$ is the change in the mean oxygen partial pressure in the system being studied and $D$ is the administered dose. ROD occurs whenever the dose delivery time is smaller than the characteristic time scale needed to restore oxygen levels depleted by radiochemical reactions [13]; *in vivo*, this is the time (~1 s) needed for oxygen to diffuse out of capillaries into surrounding tissue and dose must be delivered in $\lesssim$ 1 s to observe these effects [17]. In preclinical *in vivo* experiments, the normal tissue toxicity-reducing effects of FLASH have been observed after single-fraction doses between 10 and 50 Gy [18-21]. Based on measured $G$ values quoted above, the lower end of this range is expected to give rise to a reduction in mean tissue oxygen partial pressure by 1-4 mmHg, consistent with measured values by Cao and colleagues, who reported FLASH-induced reductions of ~1 and 2 mmHg in tumours and normal tissues, respectively, after a single 20 Gy dose [16]. These values are much smaller than the magnitude estimated previously ($\gtrsim$ 10-40 mmHg) that would naïvely be needed to significantly alter the OER in normal tissue. This led Pratx and colleagues to propose that normal tissue hypoxic stem cell "niches" ($\lesssim$ 10 mmHg) are responsible for the radiation response of normal tissues [2, 17], since such cells would be expected to be in the range of oxygen levels where changes induced during FLASH would significantly influence cell survival.

This argument hints at an important point: even in nominally well-perfused normal tissues, the vascular network is irregular [22-26], and while more efficient at delivering oxygen than the typical tumour vasculature [27], oxygen distributions still exhibit a substantial degree of heterogeneity [25, 28-30]. Consequently, in well-oxygenated normal tissues (mean partial pressures $\bar{p} \gtrsim$ 20 mmHg), there may be hypoxic regions where the local partial pressure $p \lesssim$ 10 mmHg and hence, FLASH susceptible. *Even if these are proportionately small compared to the normoxic regions, radiation-induced cell survival varies exponentially with the effective dose D·OER, and the response of tissue to radiotherapy will be determined by these most hypoxic cells.*

The heterogeneity of oxygen partial pressures within tissues is thus a potentially important factor in determining the impact of FLASH. In this work, we simulated its effect by solving the oxygen reaction-diffusion equation [31, 32] including the effect of ROD [2, 4] over domains containing capillaries (oxygen sources) distributed randomly and on the vertices of a square lattice, thereby simulating two extremes of geometric irregularity (and hence, oxygen heterogeneity). The resulting oxygenation maps were used to calculate target cell survival in normal and tumour tissues using the linear quadratic model, modified to account for oxygen sensitizing effects. These simulations were run with and without ROD, allowing us to calculate the differential cell survival between FLASH and conventional dose rate radiotherapy (herein referred to simply as "conventional radiotherapy"). Our work thus builds on the calculations of Pratx *et al.* [2] and Rothwell *et al.* [4], who carried out related simulations for a single cylindrical capillary ("Krogh geometry").

To connect these cell survival results with empirical preclinical studies showing reduced normal tissue toxicities and apparent iso-tumour control [18, 33, 34] using FLASH, we calculated the ratio of the FLASH and conventional radiotherapy doses needed to achieve iso-cell survival for normal and tumour tissues. These were compared with empirical dose-modifying factors for iso-toxicity, which have been measured in several tissues, including lung, brain [19, 33], and skin [35]. The limitations of equating cell survival with toxicity are considered in the Discussion.

## 2. Methods

*2.1 Cell survival analysis*

The differential impact of FLASH and conventional radiotherapy was quantified by estimated normal and tumour tissue cell survival fractions (SF's). SF's were calculated using the linear-quadratic (LQ) model for a single fraction of radiotherapy (following most preclinical experimental FLASH regimens), modified to account for OER effects by convolving the LQ expression for survival fraction with the time-dependent distribution $f(p,t)$ of oxygen partial pressure values $p$ within the tissue:

$$\text{SF} = \frac{1}{T}\int_0^T dt \int_0^\infty dp\, f(p,t) e^{-\alpha \cdot \text{BED}(p,D)} \tag{2}$$

Here,

$$\text{BED}(p,D) = D \cdot \frac{\text{OER}(p)}{\text{OER}_{\max}}\left(1 + \frac{D \cdot \text{OER}(p)}{(\alpha/\beta) \cdot \text{OER}_{\max}}\right) \tag{3}$$

is the biologically effective dose corresponding to a dose $D$, modified by the oxygen partial pressure-dependent OER. Equation (2) generalizes a well-known expression for the oxygen-dependent linear quadratic formula [8, 32] to the case where the oxygen distribution $f(p,t)$ is time-dependent, as happens during FLASH radiotherapy, here taken to be delivered between a time $t = 0$ and $t = T$.

The partial pressure dependence of the oxygen enhancement ratio was approximated as [8, 36, 37]

$$\text{OER}(p) = \frac{p \cdot \text{OER}_{\max} + K_m}{p + K_m} \tag{4}$$

where $\text{OER}_{\max} = 3$ is the maximum OER value, achieved when $p \gg K_m$. The partial pressure $K_m$ at which OER achieves half its maximum value is estimated to lie between 1.9 and 3.28 mmHg [8, 37]. α and β in Equations (2) and (3) quantify the linear and quadratic dependencies of radiosensitivity with respect to dose; values for these parameters and all others used in the calculation of SF are shown in Table 1.

It was assumed that Equation (2) applied equally well to normal and tumour tissues, with different values of α and β. For tumours, the ratio α/β was chosen to be 10 Gy, while for normal tissues, it was taken to be 3 Gy, standard values for early- and late-responding tissues [38]. Because the values of α under oxic conditions are highly variable between tumour [39-41] and normal [41] tissues, we simulated normal and tumour cell survival using a range of values: 0.2, 0.3, and 0.4 Gy$^{-1}$, consistent with empirical *in vitro* values. Equation (2) also assumed that the density of "target" cells (e.g., tumour clonogens) is uniform with respect to oxygenation.

The heterogeneity of oxygen partial pressures within a tissue was quantified by the distribution $f(p,t)$, which represents the probability that a given point in space will have an oxygen partial pressure $p$, a time $t$ after the start of FLASH. As noted in the Introduction, even though the mean oxygen partial pressure $\bar{p} = \int_0^\infty dp\, p f(p, 0)$ before FLASH is delivered is sufficiently large in normal tissues that $\text{OER}(\bar{p}) \sim \text{OER}_{\max}$, any irregularities in the spatial distributions of capillaries means that there may be regions at low partial pressure, $f(p{\sim}0, 0) \neq 0$. These radioresistant regions, for which OER ≈ 1, will ultimately determine SF owing to its exponential dependence on $D \cdot \text{OER}(p)$ in Equation (2). In what follows, we will suppress the time variable $t$ when discussing the pre-FLASH equilibrium oxygen partial pressure histogram: $f(p) \equiv f(p, t \leq 0)$.

*2.2 Oxygen partial pressure distributions*

Oxygen partial pressure distributions $f(p,t)$'s were calculated by solving the oxygen reaction-diffusion equation

$$\frac{\partial p(\vec{r},t)}{\partial t} = D_{O_2} \vec{\nabla}^2 p(\vec{r}) - \frac{c_{\max} \cdot p(\vec{r},t)}{k + p(\vec{r},t)} - \frac{(D/T) \cdot G_0 \cdot p(\vec{r},t)}{k_{\text{ROD}} + p(\vec{r},t)} \quad (5)$$

for the oxygen partial pressure $p(\vec{r},t)$ at position $\vec{r}$ on a two-dimensional domain, subject to

$$p(\vec{r},t) = p_c \text{ for } \vec{r} \text{ on the surface of capillaries (radius } r_c),$$

with $p_c$ the capillary oxygen partial pressure. Details of the two capillary architecture models are described below.

Equation (5) combines the effects of oxygen diffusion (diffusivity $D_{O_2}$) with metabolism (the second term on the right-hand side) and radiolytic oxygen depletion (third term on the right-hand side). Oxygen metabolism is the same as the Michaelis-Menten form used by Daşu and collaborators [42] and Petit and colleagues [32], with $c_{\max}$ the maximum metabolic rate and $k$ = 2.5 mmHg. The radiolytic oxygen depletion (ROD) term was assumed to similarly obey Michaelis-Menten kinetics; it reduces to the form used by Pratx and collaborators [2] in the limit that $k_{\text{ROD}} \to 0$. *In vitro* studies have discerned no deviation from a constant ROD as a function of oxygen partial pressure down to several mmHg [13-15], and yet ROD must vanish as the partial pressure does; correspondingly, we chose a small value for the partial pressure $k_{\text{ROD}} = 1$ mmHg at which the ROD is half its maximum value. We solved Equation (5) with and without the ROD term to simulate the effect of FLASH on oxygen distributions.

Following the approach of Weiss and others [2, 13, 43], Equation (5) assumes that ROD is simply proportional to the dose rate $D/T$, where $D$ is the dose delivered in a time $T$, and $G_0$ is the constant of proportionality. More generally, the kinetics of the radiochemical reactions responsible for ROD and oxygen fixation of radiation-induced DNA damage are properly described by second-order reaction equations [4, 44]. As shown by Ling, however, the results of calculations using these second-order reaction equations approach those using the simpler kinetics in Equation (5) for radiation pulses lasting longer than the time scales describing oxygen fixation ($\tau_{\text{fix}}$) and ROD kinetics ($\tau_{\text{ROD}}$), both on the order of several microseconds ($\mu$s) [44, 45]. (At the same time, as noted in the Introduction, the pulse duration must be smaller than the time needed for oxygen to diffuse away from capillaries for there to be substantial ROD.) For pulsed irradiation, we further require that the time between pulses is much longer than $\tau_{\text{fix}}, \tau_{\text{ROD}}$. Most recent experiments, involving either pulsed or continuous irradiation well-satisfy all these conditions (see Table 1 in the review paper by Wilson and colleagues [34]).

It was further assumed that the capillary oxygen partial pressure $p_c$ was unaffected by FLASH, based on the fact that the hemoglobin oxygen dissociation rate ~20 s$^{-1}$ [46] is much faster than typical values of $1/T$. We thus expect that ROD in blood is immediately compensated for by hemoglobin giving up oxygen to maintain a near-constant partial pressure.

Random two-dimensional capillary domains have been used previously to simulate the often-dysfunctional tumour vasculature [32, 42]. In contrast, normal tissues are often assumed to have a well-organized vasculature, optimized to meet metabolic demands, and modelled as a regular array of non-interacting cylinders (Krogh geometry) [22], including in FLASH studies [2, 4]. There is a growing appreciation, however, that the normal tissue vasculature exhibits a substantial degree of irregularity, and that this irregularity may be important for understanding tissue physiology [23, 25]. The hallmarks of such irregularity are regions of hypoxia and, related to this, a long tail in the distribution $n$(DNC) of the minimum distances between sampled points in the tissue extravascular space and the nearest capillary, i.e., the "distance-to-nearest-capillary" (DNC). This distribution as well as the closely-related minimum intercapillary distance distribution have been studied extensively in both normal [24-26, 28, 47-50] and tumour [24, 48] tissues and both distributions indeed exhibit such a tail at long distances ($> 50 \mu$m), indicating the likely presence of small regions of poorly oxygenated regions, even in healthy normal tissues [25, 28]. This has been confirmed in measurements of normal tissue oxygen partial pressure distributions, which show the presence of anoxic regions [$f(p\sim 0) \neq 0$] even in nominally well-oxygenated tissues such as liver, brain, and spinal muscle [30].

We calculated oxygen partial pressure distributions $f(p, t)$ as well as DNC histograms for both randomly distributed capillaries and capillaries distributed on the vertices of a square lattice to represent the two extremes of vascular irregularity used in the literature. Based on the similarity between $f(p)$ and DNC histograms calculated using the random capillary model and empirical tumour and normal tissue results, we argue in the Discussion that the random capillary model is a better representation of the vasculature in both these tissue types, although results are presented for both. We emphasize that we are not asserting that the normal tissue vasculature is random, only that it exhibits sufficient irregularity that the predicted oxygen partial pressure distributions and architectural metrics arising from a random capillary model better match with empirical results than e.g., a model representing a regular array of capillaries.

Parameter values used in our simulations are shown in Table 1. Of note, simulations were carried out using single-fraction doses of 20 Gy delivered in $T = 0.4$ s, two representative values of ROD, $G_0 = 0.2$ and 0.4 mmHg/Gy, a capillary oxygen partial pressure $p_c = 40$ mmHg [4], and maximum oxygen metabolic rates $c_{\max}$ of 5, 15, and 40 mmHg/s for both normal and tumour tissues. Justification for these choices is given below.

Recent *in vivo* preclinical experiments have investigated the effects of FLASH irradiation using single-fraction doses spanning 10 to 30 Gy [16, 20, 21, 51]; we used 20 Gy as a representative value, also the value used in the experiments of Cao *et al.* [16], the results of which are most comparable to our own. Assuming $T = 0.4$ s gives a dose rate of 50 Gy/s, which is a typical dose rate at which normal tissue sparing effects have been observed [34].

We use the notation $G_0$ to distinguish the rate of ROD that enters the oxygen reaction-diffusion equation, Equation (5), from the empirically observed values $G$, defined in Equation (1). As noted by Pratx and Kapp, the two values only coincide when oxygen diffusion and metabolism are ignored [2]. We use $G_0 = 0.2$ and 0.4 mmHg/Gy, approximately encompassing the range of measured values *in vitro* [13-15], where oxygen distributions are fairly uniform spatially and hence, diffusion is less relevant.

The choice of $p_c = 40$ mmHg corresponds to the expected venous oxygen tension [10, 52]. Although partial pressures drop across capillaries from ~ 90-100 mmHg in arteries to 35-45 mmHg in veins [10, 52], radiation response will depend on the least well-oxygenated components of the tissue, i.e., near the venule ends of capillaries. In the Discussion, we discuss the impact of this assumption on our results in some detail, arguing that alternative choices of $p_c$ (or any other of the values used in our simulations) do not change the conceptual results of our study. This also includes the maximum rates of oxygen metabolism $c_{\max}$, assumed here to be 5, 15, and 40 mmHg/s. The value 15 mmHg/s is a classic value for tumours [32, 42, 53, 54], based on measurements of human tumours by Warburg [53]. Individual tumours exhibit a range of metabolisms, however [55]. Similarly, normal tissues exhibit a range of metabolic activities, with liver and kidney generally exhibiting a higher metabolic rate than the average tumour, brain and intestines exhibiting comparable levels, and resting skeletal muscle exhibiting lower levels (see Ref. [30] as well as Table SM1 in the Supplementary Materials). Further discussion of tumour and normal tissue metabolic levels is given in the Supplementary Materials Section 1, where we argue that 5-40 mmHg/s encompasses an appropriate range for most normal tissues.

For a specified value of the areal capillary density $n_c$ ($= N_c/l^2$, where $N_c$ is the number of capillaries and $l = 1$ mm is the size of the simulation domain), Equation (5) was solved using a finite-element technique, having randomly distributed the capillaries over the domain or having placed them on the vertices of a regular square lattice with spacing $\approx l/\sqrt{N_c}$. Zero-flux Neumann boundary conditions were imposed on the partial pressure $p(\vec{r}, t)$ along the edges of the domain boundaries. For each capillary configuration, Equation (5) was solved for conventional RT (approximated as $(D/T) \cdot G_0 = 0$; from here on, we simply indicate conventional RT as $G_0 = 0$) and for FLASH RT ($G_0 = 0.2$ and 0.4 mmHg/Gy). The equilibrium (static) conventional RT solution was used as the initial condition $p(\vec{r}, 0)$ for the corresponding FLASH calculation, carried out for $0 \leq t \leq T$. Figure 1 shows the results of one such calculation for a random capillary array. Following the approach of Ref. [32], for the random capillary arrays, the calculation was repeated at least 100 times for different random capillary placements but fixed $n_c$, to ensure adequate statistics in the calculation of SF [Equation (2)]. For calculations involving the lattice capillary array, only a single calculation was needed for each $n_c$. At discrete timepoints $t = 0, \Delta t, 2\Delta t, \ldots, T$ separated by the increment $\Delta t$, the resulting spatial oxygen partial pressure maps $p(\vec{r}, t)$ were sampled over the 1mm × 1mm spatial domains, excluding the outermost regions < 100 $\mu$m from the boundaries, to generate oxygen partial pressure distributions $f(p, t)$ for each value of $n_c$ for conventional and FLASH radiotherapy. This exclusion was done to avoid finite-size effects arising from the zero-flux boundary conditions. For each simulation and each timepoint, 900 × 900 points were thus sampled over the domains, corresponding to one partial pressure data point per $\mu m^2$. These datasets were combined to generate a total dataset of partial pressure values for each $n_c$ and timepoint of size $N_s \geq 8.1 \times 10^7$ (= 900 × 900 × 100). Values of $n_c$ spanned 8 mm$^{-2}$ to 637 mm$^{-2}$. Examples distributions are shown in Figure 2 for three representative values of $n_c$ for $t = 0$ and $t = T$. The sampled partial pressure datasets were used to calculate SF, approximating the integral as a sum [Equation (6)] over the number $N_s$ of sampled points and a trapezoidal time integration,

$$\text{SF} \cong \frac{1}{N_s T} \sum_{j=1}^{N_t} \sum_{i=1}^{N_s} \left( \frac{e^{-\alpha \cdot \text{BED}(p_i(t_{j-1}), D)} + e^{-\alpha \cdot \text{BED}(p_i(t_j), D)}}{2} \right) \Delta t, \quad (6)$$

where $p_i(t_j)$ is the $i$th element of the sampled histogram dataset at timepoint $t_j$. $N_t \equiv T/\Delta t$ is the number of timepoints used in the integration. Because the spatial datasets were large, we used only $N_t = 5$ timepoints. At each point $\vec{r}$, $p(\vec{r}, t)$ decays smoothly with time and the numerical error associated with this relatively coarse integration step size is small. Survival fraction curves were computed as functions of the mean tissue partial pressure without ROD (equivalently, pre-FLASH), similarly discretized as shown in Equation (7):

$$\bar{p} = \int_0^\infty dp\, p f(p) \cong \frac{1}{N_s} \sum_{i=1}^{N_s} p_i(0). \quad (7)$$

Here, $p_i(0)$ is the $i$th element of the sampled histogram dataset without ROD (i.e., $G_0 = 0$). We also calculated the change

$$\Delta \bar{p} \equiv \frac{1}{N_s} \sum_{i=1}^{N_s} [p_i(T) - p_i(0)]. \quad (8)$$

in mean partial pressure over the course of FLASH; i.e., the difference in partial pressure between the end ($t = T$) and beginning ($t = 0$) of FLASH.

*2.3 Dose modifying factor*

To connect the SF results with empirical normal tissue toxicity and tumour control results, we defined the dose modifying factor (DMF) as

$$\text{DMF} \equiv \frac{20 \text{ Gy}}{D_{\text{conv, iso-SF}}}, \quad (9)$$

where $D_{\text{conv, iso-SF}}$ is the single-fraction dose of conventional radiotherapy needed to achieve the same survival fraction ("iso-SF") as 20 Gy of FLASH. Having calculated the SF for 20 Gy and the FLASH oxygen distribution for a given capillary density using Equation (6), Equation (6) was then solved iteratively using the corresponding conventional dose-rate oxygen distribution to find the dose $D_{\text{conv, iso-SF}}$ that resulted in the same SF. The DMF was calculated for both tumour ($\alpha/\beta = 10$ Gy) and normal tissues ($\alpha/\beta = 3$ Gy). Results are shown for the representative metabolic rate $c_{\max} = 15$ mmHg/s and both simulated rates of ROD.

## 3. Results

*3.1 FLASH-induced reduction in oxygen partial pressures*

FLASH lowered mean tissue oxygen partial pressures by an amount dependent on the initial mean oxygen partial pressure, the value of $G_0$, and the metabolic rate $c_{max}$; see Figure 3. Tissues with low mean partial pressure $\bar{p}$ values were largely refractory to FLASH, with $\Delta\bar{p}$ approaching zero as $\bar{p}$ did. With increasing partial pressure, $\Delta\bar{p}$ grew, reaching a maximum when $\bar{p} \gtrsim 20$ mmHg. For the random capillary array model and intermediate metabolic rate $c_{max} = 15$ mmHg/s, maximum $\Delta\bar{p}$ values were ~ 3 mmHg for $G_0 = 0.2$ mmHg/Gy and ~ 7 mmHg for $G_0 = 0.4$ mmHg/Gy, respectively. For the simulated dose of 20 Gy used in our calculations, these correspond to the maximum ROD rates $G$ slightly below $G_0$. As can be seen in Figure 3, $\Delta\bar{p}$ decreased with increasing metabolic rate, since the magnitude of the ROD effect depends on the relative sizes of $(D/T) \cdot G_0$ and $c_{max}$; see Equation (5). Results for square lattice capillary configurations were quantitatively similar, and only results for the random capillary model are shown.

*3.2 Impact of FLASH on tumour and normal tissue cell survival*

For both models of capillary architecture studied, FLASH led to a pronounced increase in cell survival fraction (SF), depending on the mean tissue oxygen partial pressure $\bar{p}$ values. Example plots of SF's versus $\bar{p}$ for $\alpha = 0.3$ Gy$^{-1}$ and $\alpha/\beta = 3$ Gy are shown in Figure 4 for the three metabolic rates studied ($c_{max}$ = 5, 15, 40 mmHg/s), a dose of 20 Gy, and the random (top) and lattice (bottom) capillary architectures. SF curves for the other simulated values of α and β were qualitatively similar and are shown in the Supplementary Materials Section 2.

At low $\bar{p}$ values (hypoxic tissue), tissues were again found to be refractory to the impact of FLASH, and the SF curves for conventional and FLASH radiation coincided in the limit $\bar{p} \to 0$. For tumours ($\alpha = 0.3$ Gy$^{-1}$ and $\alpha/\beta = 10$ Gy), the difference in SF's between FLASH and conventional radiotherapy remained $\lesssim 1$ order-of-magnitude for $\bar{p} \lesssim 10$ mmHg (the expected range for most tumours), using $c_{max} = 15$ mmHg/s; see Figure SM1 in the Supplementary Materials.

With increasing $\bar{p}$, the SF curves separated; for the lattice capillary geometry, a maximum separation was achieved for $\bar{p}$ ~5-20 mmHg (depending on parameter values), while for the random capillary model, the curves separated monotonically with increasing $\bar{p}$. For the random capillary model, the separation was substantial, with normal tissue cell survival increasing by as much as 10 orders of magnitude, depending on the relative magnitudes of $G_0$ and $c_{max}$ (Figure 4).

*3.3 Dose modifying factors*

This differential cell-killing manifested itself in tumour and normal tissue dose-modifying factors ≥ 1, rising from 1 at $\bar{p} = 0$ to maximum values approaching 1.7 (for $c_{max} = 15$ mmHg/s and $G_0 = 0.4$ mmHg/Gy); see Figure 5. Similar to the survival fraction results, although the magnitude of the maximum DMF was comparable for the random and lattice capillary models (not shown), the maximum DMF for the lattice model was achieved at relatively low mean partial pressure values, $\bar{p}$ ~ 5 −10 mmHg, before decreasing rapidly. In contrast, for the random capillary model, DMF continued to grow with increasing $\bar{p}$, achieving a maximum for $\bar{p} \gtrsim 20$ mmHg, the expected range of oxygen tensions for normal tissues. To the extent that tumours have lower $\bar{p}$ than normal tissues, the impact of FLASH was thus greater for normal tissues than for tumours. Figure 5 shows representative DMF curves for tumour ($\alpha/\beta$ = 10 Gy) and normal ($\alpha/\beta$ = 3 Gy) tissues using $\alpha$ = 0.3 Gy$^{-1}$, $c_{max}$ = 15 mmHg/s, and different values of the rate $G_0$ of ROD. $c_{max}$ = 15 mmHg/s was assumed here to represent both tumour oxygen metabolism as well as normal tissue at an intermediate metabolic rate; see Supplementary Materials Table SM1.

## 4. Discussion

By depleting tissues of oxygen, FLASH radiotherapy was found to increase cell survival because of the oxygen enhancement ratio (OER) effect. The difference in cell survival between FLASH and conventional radiotherapy grew by orders of magnitudes with increasing tissue oxygenation, providing a possible explanation of the observed reductions in normal tissue toxicity and apparent iso-tumour control [18, 33, 34], since tumours are often found at lower oxygen levels than normal tissues, with mean oxygen tensions typically below 10 mmHg [56-61]. For completely anoxic tissue (i.e., oxygen partial pressure exactly zero), FLASH cannot further deplete oxygen from tissue and the survival curves for FLASH and conventional radiotherapy coincide.

The substantial impact of FLASH on cell survival in well-oxygenated tissue is the central result of this manuscript, contradicting the naïve expectations that oxygen sensitization effects based on *mean* tissue partial pressures should be small [2], and that oxygen must be depleted completely for the ROD mechanism to contribute significantly to toxicity reductions [11, 12]. The relative increase in radioresistance with increasing oxygenation is a result of the heterogeneity of the oxygen distribution in tissues. A single dose of 20 Gy kills all well-oxygenated cells (SF~$10^{-20}$ for OER = OER$_{max}$ using e.g., $\alpha$ = 0.3 Gy$^{-1}$ and $\alpha/\beta$ = 3 Gy) and hence, the survival fraction is essentially the fraction $f(p{\sim}0)$ of maximally radioresistant cells at ultra-low oxygen levels. FLASH increases cell survival by increasing the number of cells in this population. Because $f(p{\sim}0)$ is small for well-oxygenated tissues (see e.g., the middle and right panels of Figure 2), the increase in $f(p{\sim}0, t > 0)$ during FLASH is proportionately greater for such tissues and the relative cell survival—FLASH versus conventional radiotherapy—likewise grows with increasing mean partial pressure. The exponential nature of cell survival after radiation means that even a very small fraction of hypoxic cells (e.g., less than 0.5% of the non-FLASH irradiated tissue represented in the right panel of Figure 2 is estimated to have partial pressures less than 5 mmHg) will have a large impact.

To different extents, this behaviour arose in both the capillary geometry models that we used, the random capillary model as well as the lattice model. However, the effect was far more robust using the random capillary model. For the lattice model, the window of mean tissue partial pressures where FLASH was predicted to result in a substantial sparing effect was substantially reduced compared to the random capillary geometry since the lattice geometry eliminates the possibility of any tissue being overly distant from capillaries. On geometric grounds, for the lattice geometry, $f(p{\sim}0, t)$ will be nonzero only when the maximum distance-to-nearest capillary (i.e., the point in the middle of the lattice unit cell) is greater than the critical distance $R_c(\tilde{c}_{max}) \equiv \sqrt{4D_{O_2}p_c/\tilde{c}_{max}}$ past a single capillary beyond which the oxygen tension vanishes [53]. Equivalently, when the capillary density satisfies

$$n_c \lesssim \frac{1}{2R_c^2(\tilde{c}_{max})} \equiv \frac{\tilde{c}_{max}}{8D_{O_2}p_c}. \tag{10}$$

Here, $\tilde{c}_{\max} = c_{\max} + G_0(D/T)$ is the approximate effective metabolic rate, accounting for the effects of FLASH [In Equation (5), FLASH is formally equivalent to a change in oxygen metabolism in the limit where the Michaelis-Menten parameters are identical, $k_{\text{ROD}} = k$]. For larger capillary densities (equivalently, larger mean tissue partial pressures, $\bar{p}$), $f(p\sim 0, t)$ will be zero for all $t \leq T$, and the effect of FLASH on cell survival vanishes:

$$\text{lattice geometry:} \quad \lim_{n_c \gg 1/2R_c^2} \text{SF}_{\text{FLASH}} \to \text{SF}_{\text{conv}} \qquad (11)$$

In contrast, for the random capillary model, $f(p\sim 0, t)$ decays exponentially at large mean tissue partial pressure, but remains nonzero, yielding an exponential separation of the survival curves with increasing tissue oxygenation ($n_c$):

$$\text{random capillary geometry:} \quad \frac{\text{SF}_{\text{FLASH}}}{\text{SF}_{\text{conv}}} \sim \exp\left[\eta \frac{n_c D_{O_2} p_c}{c_{\max}} \left(1 - \frac{c_{\max}}{\tilde{c}_{\max}}\right)\right] \qquad (12)$$

Here, $\eta$ is a constant of order unity; see the Supplementary Materials Section 3 for a derivation of this result. These respective behaviours—the absence of anoxic regions in the lattice model above a critical capillary density and the convergence of the survival curves for the lattice model and exponential separation for the random capillary model—are apparent in Figures 2 and 4.

The question of which model—random capillary or lattice array—better describes the actual normal and tumour vascular architectures is thus an important one, insofar as only the random capillary model predicts a *robust* increase in the impact of FLASH with increasing tissue oxygenation: as evident from Equation (12), the choice of parameter values $n_c$, $D_{O_2}$, $p_c$, $c_{\max}$, and $G_0$, used in our simulations will affect the rate at which the survival curves separate with increasing oxygenation, but not the *existence* of such a therapeutic advantage. In contrast, for a regular lattice capillary array, unless Equation (10) is satisfied, the advantage will not exist. Note that to satisfy the metabolic demands of the majority of a given normal tissue, one would generally expect that $n_c \gtrsim 1/2R_c^2$ and hence, Equation (10) may not be satisfied for normal tissues, with $f(p\sim 0)$ vanishing for physiologically relevant parameter values. As noted above, it is the fact that $f(p\sim 0)$ is nonzero—albeit it, very small—in the random capillary model even at high mean tissue oxygen tensions, $\bar{p} \gtrsim 20$, that produces the large separation of survival curves, with FLASH producing a significant tissue sparing effect. Empirical studies have indeed found small but nonzero $f(p\sim 0)$ even in well-perfused normal tissues such as liver and brain where $\bar{p} \gtrsim 20$ mmHg [30], in sharp contrast to the $f(p)$'s calculated using the lattice capillary model (see e.g., the lower panel in Figure 2), for which $f(p\sim 0, t)$ is identically zero for large capillary densities, consistent with Equation (10).

This behaviour supports the conclusion that, in terms of the predicted oxygen distributions at least, the random capillary model may be a better model of the normal and tumour tissue vasculature than a regular array model. Further evidence for this can be found by comparing calculated distance-to-nearest capillary (DNC) histograms. As noted in the Introduction, vessel morphometry studies have quantified this distribution and the closely-related minimum intercapillary distance distribution in normal [24-26, 28, 47-50] and tumour [24, 48] tissues. A key hallmark of vascular irregularity is a "tail" in these distributions at long distances ($\gtrsim 50$ $\mu$m), indicating the presence of small regions of poor vascularization. In Supplementary Materials Section 4, we show the results of a calculation of $n$(DNC) for a well-perfused simulated normal tissue with a random vasculature and compare it with results for tissue with a lattice of capillaries. The former is found to result in an $n$(DNC) that qualitatively matches empirical normal tissue results, while the latter fails to reproduce the characteristic tail. Hence, we conclude that the random approximation is a likely an adequate model of normal and tumour tissue vasculature to predict the impact of FLASH on both these tissues. We emphasize, however, that differences in vascular architectures amongst normal tissues (e.g., brain versus skin) will result in different normal tissue sensitivities to radiation, and a better understanding of vascular irregularity as well as oxygen heterogeneity in different tissues will help improve our assessment of the magnitude of FLASH's impact.

The fact that the dose-modifying effects of FLASH are only felt by cells at low local oxygen partial pressures led Pratx and colleagues to hypothesize that normal tissue stem cells in hypoxic "niches" are primarily responsible for the radiation response of normal tissues to radiation [17]. Our tissue oxygenation distribution simulations suggest that such hypoxic niches may be present in all tissues, even nominally well-vascularized ones. Whether or not the target cells (normal tissue stem cells in the work of Pratx *et al.*) preferentially populate low-oxygen regions or are uniformly distributed with respect to oxygenation (as assumed in this work), the high oxygen enhancement ratio of cells in these niches will protect them from radiation, characterized by the large enhancement in cell survival fraction.

The magnitude of the impact of FLASH—both on the ROD-induced change $\Delta \bar{p}$ in mean oxygen partial pressure and the change in cell survival fractions—depended on our chosen parameter values. Although a comprehensive assessment of the range of parameter values for which FLASH can have a substantial effect has been undertaken by Rothwell and colleagues, who also used a more detailed model of FLASH kinetics [4], we make two observations here. Using dimensional analysis, the impact of FLASH depends on the magnitudes of the two dimensionless parameters

$$x \equiv \frac{n_c D_{O_2} p_c}{c_{\max}} \equiv \frac{n_c R_c^2(c_{\max})}{4}, \qquad y \equiv \frac{(D/T) \cdot G_0}{c_{\max}} \qquad (13)$$

As noted previously, $x$ is expected generally to be on the order unity for normal tissues, ensuring that most of the tissue is at nonzero oxygen partial pressures (note that this does not preclude the possibility that $f(p \sim 0) \neq 0$, only that it is a small number, much less than one). For all our choices of parameter values corresponding to expected normal tissue mean partial pressure values, $\bar{p} \gtrsim 20$ mmHg, $x$ varied from $\sim 0.5$ to $\sim 3$, indicating that our parameter choices were physiologically reasonable. *Hence, irrespective of our precise choices of parameter values—$n_c$, $D_{O_2}$, $p_c$, and $c_{\max}$—the large separation between survival fraction curves shown in Figure 4 is expected to hold for normal tissues, defined here as tissue for which $\bar{p} \gtrsim 20$ mmHg, but more generally as tissue for which $x \gtrsim 1$*. This leaves $y$ as the only "free" parameter, unconstrained by physiology. It represents the ratio of the ROD-induced temporal rate of oxygen depletion and the rate of oxygen metabolism. If this quantity is small, then ROD represents a small perturbation on top of the baseline oxygen metabolism; conversely, if it is large, ROD substantially renormalizes the effective oxygen metabolism. Our smallest simulated $y$ value—corresponding to $c_{\max} = 40$ mmHg/s and $G_0 = 0.2$ mmHg/Gy— was 0.025. Use of this value gave rise to a maximum $\Delta \bar{p}$ value of $\sim 2$ mmHg (see Figure 3), yielding maximum values $G \equiv \Delta \bar{p}/D \sim 0.1$ mmHg/Gy of the empirical rate of ROD, smaller than $G_0$. Our largest value of $y$—corresponding to $c_{\max} = 5$ mmHg/s and $G_0 = 0.4$ mmHg/Gy— was 4. Use of this value gave rise to a maximum $\Delta \bar{p}$ value of $\sim 6.5$ mmHg, corresponding to $G = 0.33$ mmHg/Gy, again slightly below $G_0$. In comparison, empirical rates of ROD span 0.1-0.42 mmHg/Gy [13-16]. Hence, our choice of values for $c_{\max}$ and $G_0$ largely encompassed observed rates of ROD.

The fact that $\Delta\bar{p}$ was predicted to vary with respect to $\bar{p}$ (Figure 3) is a consequence of oxygen heterogeneity and is an important differentiating feature between *in vitro* and *in vivo* experiments, since oxygen should be much more homogeneously distributed in the former. In several *in vitro* experiments, $\Delta\bar{p}$ was found to be nearly independent of the pre-FLASH mean oxygen partial pressure $\bar{p}$ down to several mmHg [13-15]. In contrast, in a recent *in vivo* experiment, Cao *et al.* found $\Delta\bar{p} \sim 1$ mmHg for tumours (pre-FLASH $\bar{p} \sim 10 - 14$ mmHg) but $\Delta\bar{p} \sim 2.3$ mmHg for normal tissues (pre-FLASH $\bar{p} \sim 20 - 35$ mmHg) [16], for 20 Gy. This is close to the behavior shown in Figure 3 for e.g., $G_0$ = 0.2 mmHg/Gy and $c_{\max}$ = 15 mmHg/s. As with our survival fraction results, this behaviour can be understood as arising from the persistence of a fraction $f(p \sim 0, t \leq T) \neq 0$ of the tissue that is refractory to FLASH, resulting in the slow increase in $\Delta\bar{p}$ with respect to $\bar{p}$ in our simulations, likely mimicking the *in vivo* situation. In contrast, the relative homogeneity of oxygen *in vitro* means that there is no such fraction for mean partial pressures above several mmHg's and the decrease in mean partial pressure is independent of $\bar{p}$.

Because of the up to 10 order-of-magnitude increase in cell survival fractions for well-oxygenated tissues ($\bar{p} \gtrsim 20$ mmHg; Figure 4) using the random capillary model, for the range of parameters studied by us, FLASH doses would need to be increased by 10%-70% to achieve the same survival fraction as conventional radiotherapy (Figure 5). This is comparable to the range of empirical iso-toxicity dose modifying factors, ~1.2-1.8, observed pre-clinically [18, 19, 33, 35, 62, 63] (see Ref. [34] for a review of DMF's). Note that some of the higher (>1.4) empirical dose-modifying factors established iso-toxicity using larger doses of FLASH radiation, 25 -75 Gy [18, 34, 35, 63], which are expected to result in greater radiolytic oxygen depletion than the 20 Gy used in this work.

The effective dose-modifying impact of FLASH radiotherapy on normal tissues should be contrasted with the apparent iso-efficacy of FLASH and conventional radiotherapy in terms of preclinical tumour volume kinetics [18, 33, 34]. Although our results predict that FLASH should have a substantially smaller impact on poorly oxygenated tissues such as tumours, the apparent absence of any effect in preclinical experiments is striking. It is likely that the effect of FLASH on tumours is multi-factorial, producing an enhancement in radioresistance via ROD, but perhaps benefitting from elevated immune response [34, 64, 65] or differences in radiochemistry as compared to normal tissues. An interesting recent proposal by Spitz and colleagues has noted that differences between levels of labile iron and metabolism between normal and tumour tissues may give rise to different levels of radiation-induced organic hydroperoxides, and hence, DNA damage [3]. More work needs to be done, however, to verify the apparent iso-efficacy for more clinically relevant doses and fractionation schedules, and for tumours exhibiting different levels of measured hypoxia.

Beyond the modelling assumptions detailed in the Methods section and the parameter sensitivity analysis presented above, a major assumption of our work concerned the relevance of linear-quadratic model cell survival calculations for normal tissue toxicity. Other computational studies have likewise considered the impact of FLASH on the OER and survival [2] as well as just the OER [2, 4]. A major potential limitation of these studies, including our own, is the assumed correlation between cell survival and normal tissue toxicity. Normal tissue toxicities are believed to be multi-factorial, possibly depending on immune response [66] and damage by radiotherapy-induced free radicals [67], in addition to radiation-induced DNA damage and subsequent cell killing. There is no clear relationship between cell survival and the probability of normal tissue toxicity. Despite this, iso-toxicity fractionation regimes have been shown to be well-described by the linear quadratic model [38], providing evidence that target cell survival is likely to play a major, if not the dominant, role in toxicity.

A recent alternative explanation for the potential large difference in cell survival curves (FLASH versus conventional) at high oxygen tensions has been proposed by Petersson and colleagues [68]. Neglecting the spatial heterogeneity of oxygen considered in the present work, they developed a novel model of ROD kinetics and found a maximum separation between mean oxygen enhancement ratios at relatively high oxygen partial pressures, $\gtrsim 20$ mmHg [68]. This is important since, as with our work, it provides a mechanism by which oxygen enhancement ratio effects can be significant even in nominally well-oxygenated tissues.

## 5. Conclusions

Modelling the heterogeneity in oxygen partial pressures that are expected to arise in tissues, we demonstrated that FLASH radiotherapy can substantially increase target cell survival as compared to conventional radiotherapy, even though the reduction in mean tissue oxygen levels was not nearly large enough to completely deplete tissue of oxygen. This effect was more pronounced for better-oxygenated normal tissues with proportionately smaller hypoxic fractions than hypoxic tumours, possibly explaining the observed differential impact of FLASH on normal tissue toxicities and tumour response. Our results highlight the importance of quantifying intra-tissue oxygen heterogeneity to fully assess the impact of FLASH on tissue radiation response.

| Parameter | Symbol | Value(s) | Reference |
|---|---|---|---|
| Oxygen diffusivity | $D_{O_2}$ | 2000 $\mu m^2/s$ | [54] |
| Maximum oxygen metabolic rate | $c_{max}$ | 5, 15, 40 mmHg/s | [42] and Supplementary Materials |
| Oxygen partial pressure at which metabolism is half its maximum | $k$ | 2.5 mmHg | [42] |
| Intrinsic rate of radiolytic oxygen depletion | $G_0$ | 0.2 and 0.4 mmHg/Gy | [13-15]. |
| Oxygen partial pressure at which ROD is half its maximum | $k_{ROD}$ | 1 mmHg | ** |
| Capillary radius | $r_c$ | 5 $\mu m$ | - |
| Capillary oxygen partial pressure (venous end) | $p_c$ | 40 mmHg | [42, 52] |
| Intrinsic radiosensitivity | α | 0.2-0.4 $Gy^{-1}$ | [39-41] |
| Alpha-beta ratio | α/β | 10 Gy for tumors; 3 Gy for normal tissues | [38] |
| Oxygen partial pressure at which the OER is half its maximum | $K_m$ | 3.28 mmHg | [8] |
| maximum OER value | $OER_{max}$ | 3 | [8] |

Table 1. Parameter definitions and values. *Range of values chosen to encompass empirically determined rates of ROD.

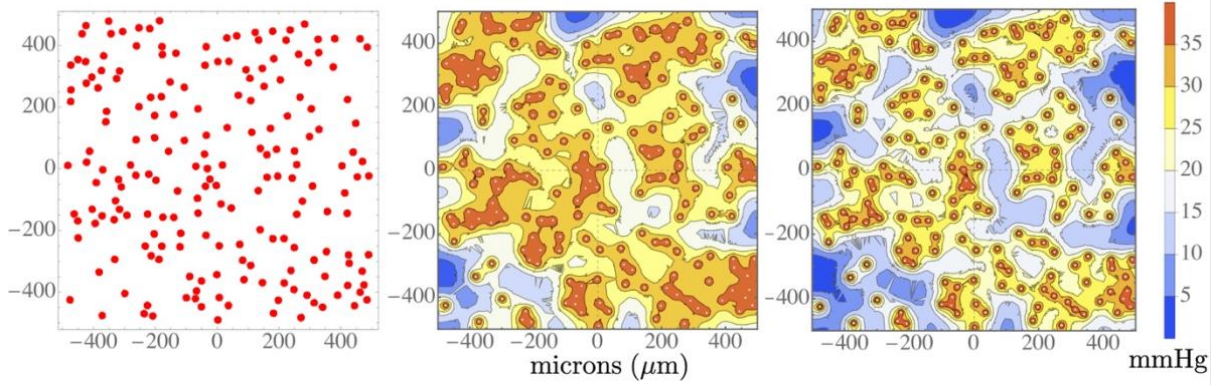

Figure 1. Simulated oxygen diffusion, metabolism, and radiolytic depletion in a 1 mm × 1 mm domain (axis values indicate $\mu$m's) for capillary oxygen partial pressure $p_c$ = 40 mmHg and maximum rate $c_{max}$ = 15 mmHg/s of oxygen metabolism. Left: Example simulation domain showing randomly distributed capillaries (corresponding to an areal density $n_c$ = 111 mm$^{-2}$; capillary radii doubled to enhance visibility). Middle: oxygen partial pressure map (mmHg) for conventional RT (equivalently, before FLASH RT) for the same capillary pattern. Right: oxygen partial pressure map for FLASH with $G_0$ = 0.4 mmHg/Gy at the end of the FLASH pulse, $t = T$ = 0.4s. Scale bar denotes oxygen partial pressures (mmHg).

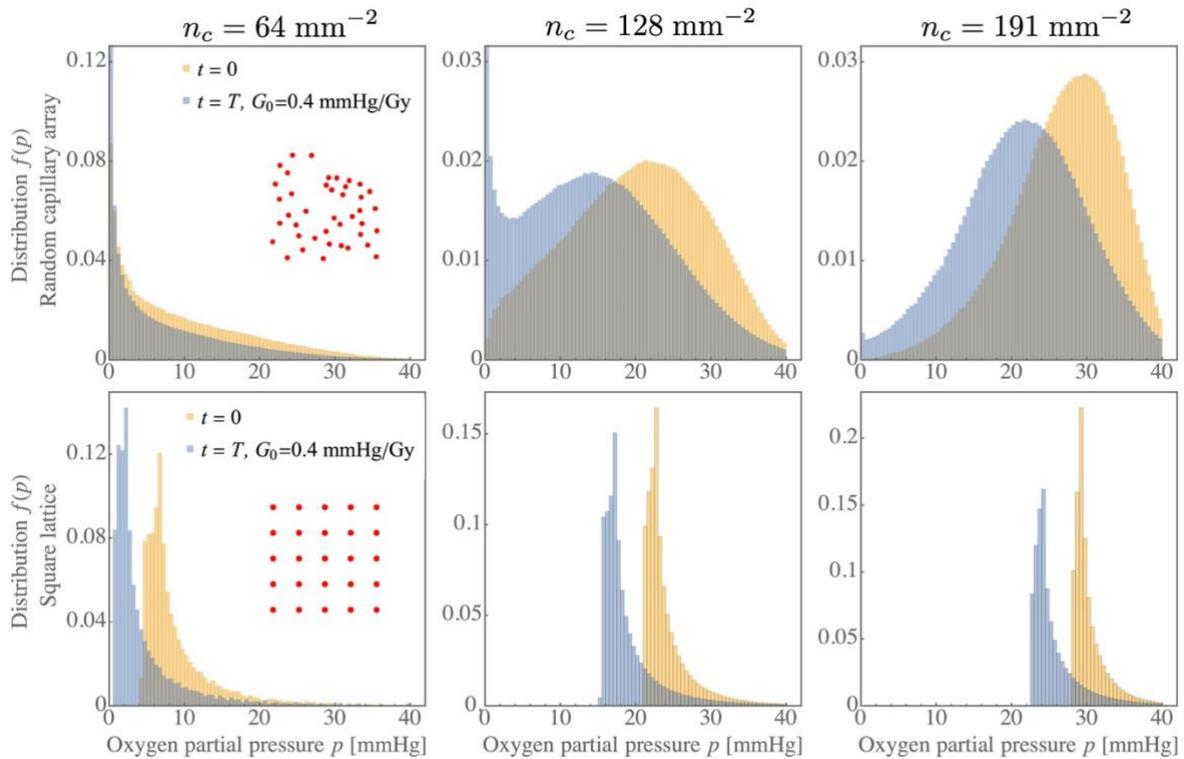

Figure 2. Impact of FLASH on oxygen distributions. Each column shows the partial pressure distributions for three representative capillary densities (indicated at top) during conventional radiotherapy (equivalently, before the beginning of FLASH, at $t$ = 0) and FLASH with $G_0$ = 0.4 mmHg/Gy at the end of the FLASH pulse ($t = T$). Simulations were carried out for $c_{max}$ = 15 mmHg/s. Top: Oxygen distributions for random capillary arrays; Bottom: Oxygen distributions for square lattice capillary arrays. The distributions for the random capillary array retain a nonzero fraction of tissue near zero oxygen partial pressure, $f(p\sim 0,t) \neq 0$, even for tissues with high mean partial pressure. In contrast, $f(p\sim 0,t)$ vanishes above a critical capillary density for the lattice array: Using Equation (10) for conventional radiotherapy, $\tilde{c}_{max} = c_{max} = 15$ mmHg/s, this is predicted to by $n_c \sim 23$ mm$^{-2}$.

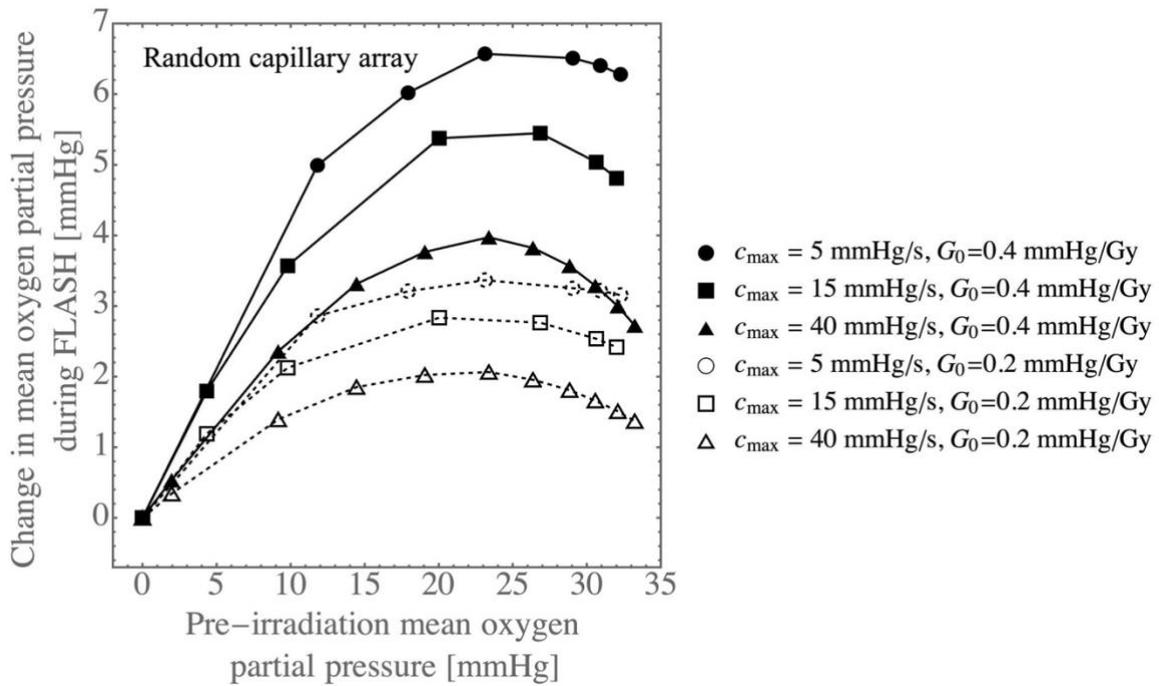

Figure 3. Change $\Delta \bar{p}$ in mean tissue oxygen partial pressure due to 20 Gy of FLASH as a function of initial mean partial pressure for the random capillary array model for different values of the oxygen metabolism rate $c_{max}$ and rate $G_0$ of radiolytic oxygen depletion. Curves for the lattice array are quantitatively similar.

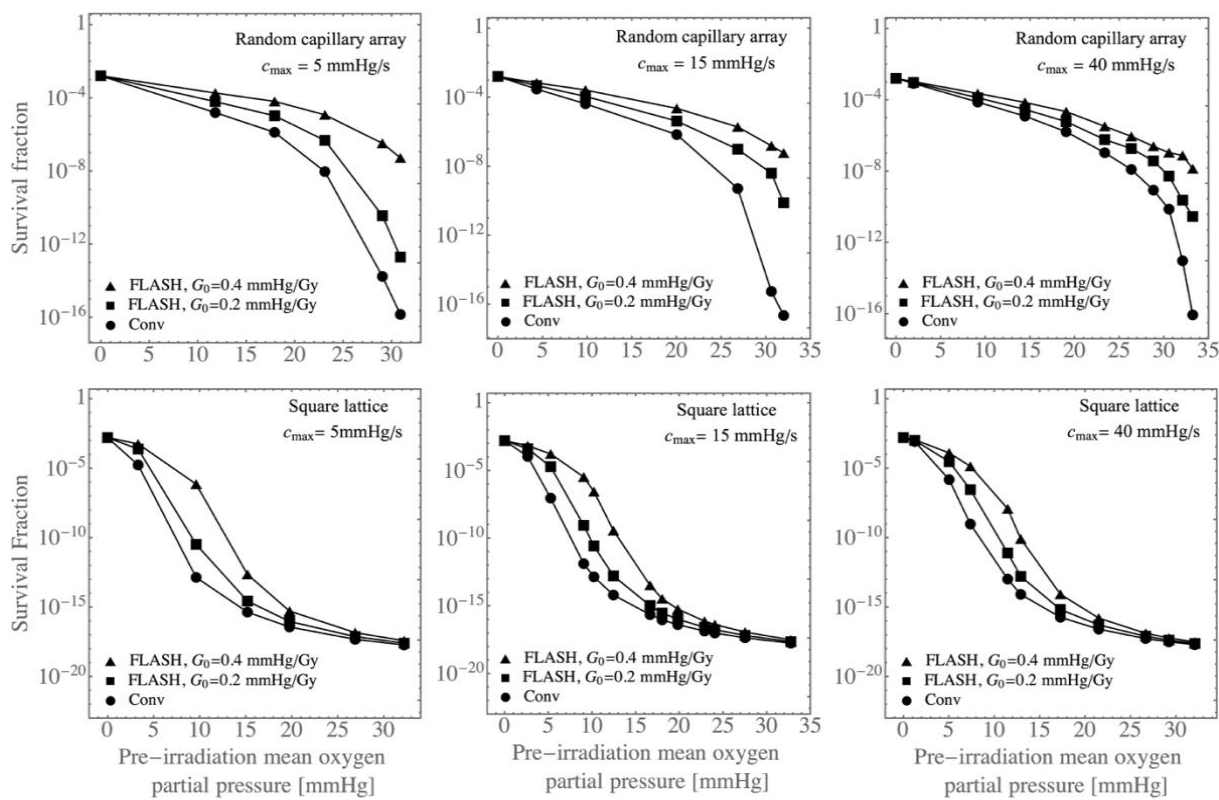

Figure 4. The impact of FLASH on cell survival depends on tissue oxygenation. Cell survival curves for a single-fraction dose of 20 Gy are shown as functions of pre-irradiation mean tissue oxygen partial pressure for different values of metabolism $c_{max}$ and rate of radiolytic oxygen depletion $G_0$ for the random (top row) and lattice (bottom row) capillary models using the representative normal tissue parameters $\alpha = 0.3$ Gy$^{-1}$ and $\alpha/\beta = 3$ Gy. Curves using other values of $\alpha$ and $\alpha/\beta$ are qualitatively similar and are shown in the Supplementary Materials Section 2.

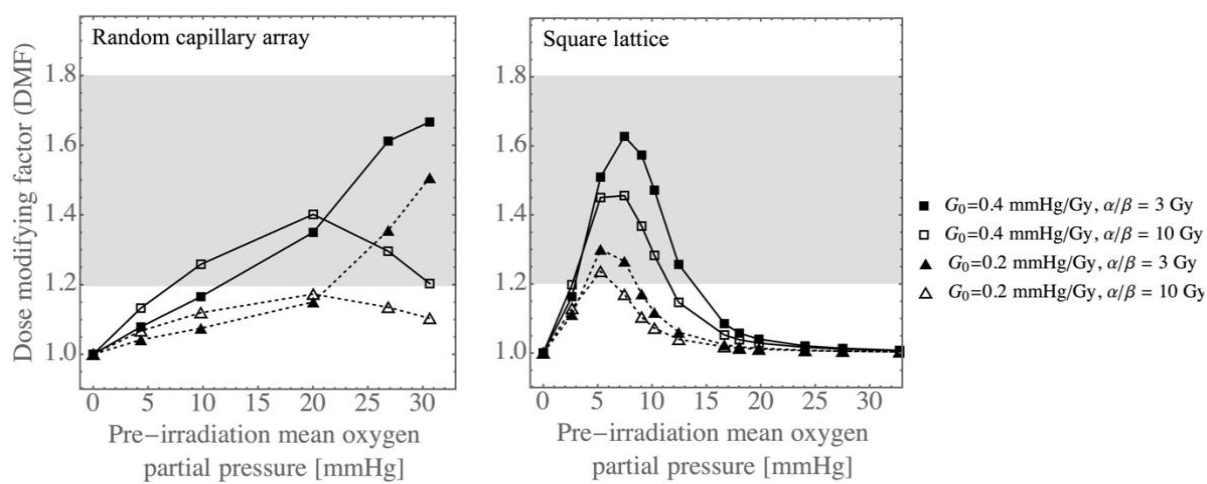

Figure 5. Iso-survival fraction dose-modifying factors (DMF's) for the random (left) and lattice (right) capillary models corresponding to a single-fraction FLASH dose of 20 Gy. Curves are shown for early- and late-responding tissues ($\alpha/\beta = 10$ and 3 Gy, respectively), $\alpha=0.3$ Gy$^{-1}$, and $c_{max} = 15$

mmHg/s. The shaded regions in indicates the range of empirical dose-modifying factors, ∼1.2-1.8 [19, 33, 35], defined in terms of normal tissue iso-toxicity.

# Modeling the impact of spatial oxygen heterogeneity on radiolytic oxygen depletion during FLASH radiotherapy: SUPPLEMENTARY MATERIALS

## 1. Rates of oxygen metabolism in normal and tumour tissues.

Theoretical studies of tumour oxygenation levels often use the value 15 mmHg/s for the maximum metabolic oxygen rate of tumours [1-3]. This result derives from an average of measurements of 13 human tumours by Warburg, determined by him to be 5.2 $\mu l_{O2}$/mg dry weight of tissue/hour [1]. To convert this figure to "wet" weight of tissue (i.e., per gram of living tissue), in their study of tumour oxygenation, Thomlinson and Gray assumed that the fraction $f$ of living tissue not comprised of water was 25%; i.e., $f = 0.25$. Thus, the metabolic rate per gram of living tissue is $\approx 22$ $\mu l_{O2}/g_{tissue}$/min (using a standard unit, see Table SM1). Thomlinson and Gray converted this value into units of mmHg/s by applying Henry's law to determine the amount of oxygen dissolved in water at 37 degrees Celsius and 1 atmosphere of pressure (760 mmHg), using a solubility of 0.024 $ml_{O2}/ml_{water}$/atm $\approx (1-f) \times 24$ $\mu l_{O2}/g_{tissue}$/760 mmHg $\approx 0.024$ $\mu l_{O2}/g_{tissue}$/mmHg. Thus, tumour metabolism was estimated to be (22 $\mu l_{O2}/g_{tissue}$/min) $\times$ (1 min/60 s) / (0.024 $\times$ $\mu l_{O2}/g_{tissue}$/mmHg) $\approx$ 15 mmHg/s.

The rate of oxygen metabolism for various normal tissues in units of $\mu l_{O2}/g_{tissue}$/min are shown in Table SM1. The corresponding rates in units of mmHg/s are also shown, using the same conversion factor applied by Thomlinson and Gray.

| Tissue | Oxygen metabolism rate | | Reference |
|---|---|---|---|
| | $\mu l_{O2}/g_{tissue}$/min | mmHg/s | |
| Human tumours | 22 | 15 | [1] |
| Human tumours | 15 | 10 | [4] |
| Brain | 33 | 23 | [5] |
| Liver | 60 | 41 | [6] |
| Intestine | 20 | 14 | [6] |
| Skeletal muscle (human quadriceps, at rest) | 8 | 5 | [7] |

Table SM1: Rates of oxygen metabolism for various tumours and normal tissues. References refer to metabolic rates in units of $\mu l_{O2}/g_{tissue}$/min; conversions to mmHg/s utilize the conversion factor employed by Thomlinson and Gray [1].

As noted by Vaupel and colleagues [8], tumours generally exhibit a level of oxygen metabolism that is intermediate between the fastest oxygen-metabolizing tissues such as liver and the slowest metabolizing tissues, such as skeletal muscle at rest. Although considerable variation exists in the metabolic rate between different tumours [4, 9], the "average" tumour also has an oxygen metabolic rate comparable to that of intestine and brain.

Based on these literature values, we use the metabolic rates $c_{max} = 5$, 15, and 40 mmHg/s in the simulations described in the main text, encompassing the expected rates for normal and tumour tissues (Table SM1).

## 2. Survival fraction curves for different values of $\alpha$ and $\alpha/\beta$.

Survival fraction curves for the random and lattice capillary model and different rates of ROD and oxygen metabolism are shown in Figure 4 in the main text using the radiobiological parameters $\alpha = 0.3$ Gy$^{-1}$ and $\alpha/\beta = 3$ Gy. Results for different parameter values are qualitatively similar. In Figure SM1, we show results for the random capillary model, metabolic rate $c_{max} = 15$ mmHg/s, and both rates of ROD studied. Note that the curves are qualitatively similar to each other, a result of the SF for $D = 20$ Gy being closely approximated by the fraction of tissue near zero oxygen partial pressure, irrespective of the values of $\alpha$ and $\alpha/\beta$ [see Equation (SM3)].

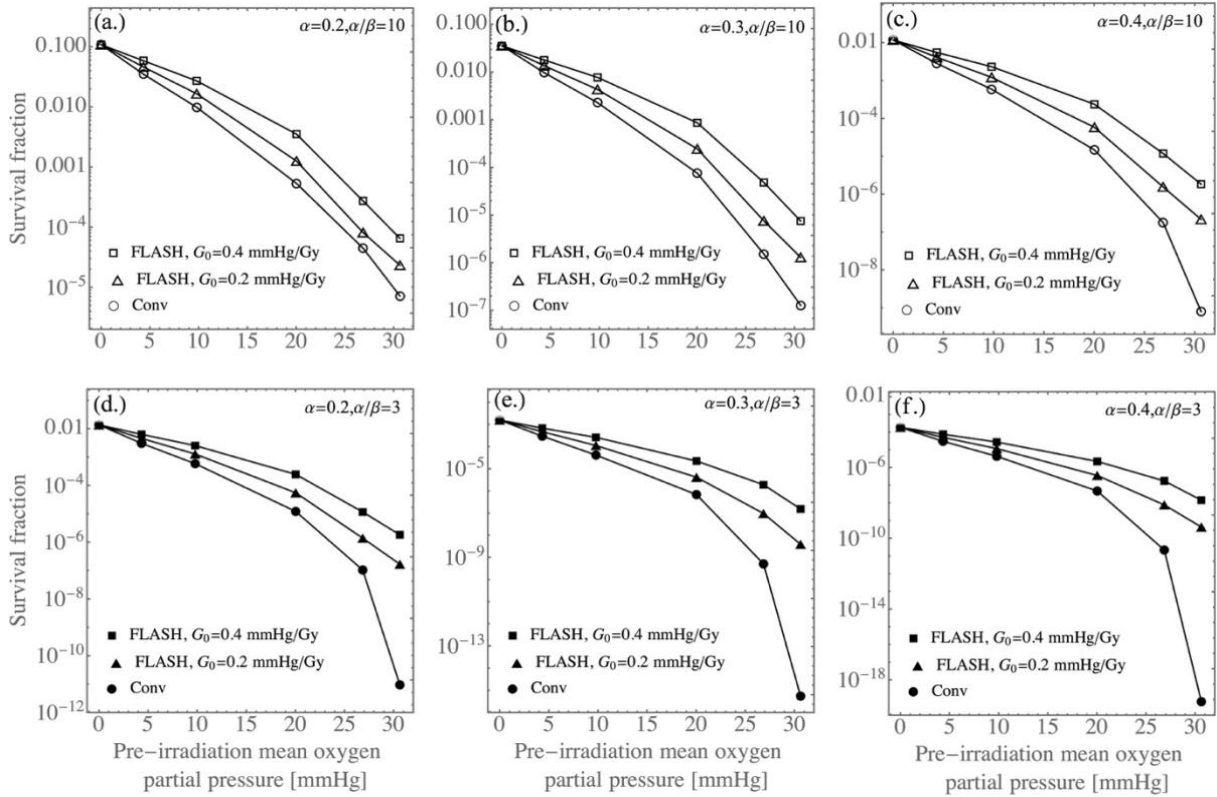

Figure SM1. The impact of FLASH on cell survival increases for increasing oxygenation. Cell survival curves as a function of pre-irradiation mean tissue oxygen partial pressure are shown for 20 Gy of conventional low-dose-rate radiotherapy and FLASH, with tissue oxygen metabolic rate $c_{max} = 15$ mmHg/s, $G_0 = 0.2$ and $0.4$ mmHg/Gy, for different values of $\alpha$ shown in the panels. The top panel [(a.)-(c.)] shows survival curves for early-responding (tumour) tissues ($\alpha/\beta = 10$ Gy); the bottom panel [(d.)-(f.)] shows curves for late-responding (normal) tissues ($\alpha/\beta = 3$ Gy).

## 3. Analytic formula for differential cell survival curves in the random capillary model

For a random distribution of capillaries in two dimensions, the fraction HF$_\Lambda$ of tissue that has an oxygen tension less than a value $\Lambda$ can be approximated from Poisson statistics:

$$\mathrm{HF}_\Lambda \sim e^{-\pi n_c r_\Lambda^2}. \tag{SM1}$$

Here,

$$r_\Lambda = \sqrt{\eta_\Lambda D_{O_2} p_c / c_{\max}} \tag{SM2}$$

is the distance from a capillary at which oxygen levels are reduced to $\Lambda$ by diffusion and metabolism, $n_c$ is the areal capillary density, $D_{O_2}$ is the diffusivity of oxygen, $p_c$ is the capillary oxygen tension, $c_{\max}$ is the rate of oxygen consumption, and $\eta_\Lambda$ is a constant of order unity that depends on the chosen threshold. For e.g., $\Lambda = 0$, $\eta_\Lambda = 4$ in two spatial dimensions [1]. Increasing $\Lambda$, $\eta_\Lambda$ will decrease from this value. The exponent in Equation (SM1) represents the expected fractional value of tissue that is within a distance $r_\Lambda$ of a capillary. $\text{HF}_\Lambda$ is thus the fraction of tissue that is more than $r_\Lambda$ from the nearest capillary and hence, hypoxic.

Choosing $\Lambda \sim$ 0-5 mmHg to coincide with the tension below which the OER changes most rapidly and further assuming that only these most hypoxic cells survive radiotherapy, the survival fraction is approximated by $\text{HF}_\Lambda$:

$$\text{SF} \approx \text{HF}_\Lambda. \tag{SM3}$$

To the extent that the Michaelis-Menten parameters $k$ and $k_{\text{ROD}}$ are not very different in magnitude, FLASH radiotherapy can further be approximated (see Equation 5 in the main text) as a reduction in metabolism,

$$c_{\max} \to c_{\max} + G_0(D/T) \equiv \tilde{c}_{\max} \tag{SM4}$$

Combining Equations (SM1)-(SM4), the ratio of FLASH to conventional survival fractions is approximated as

$$\frac{\text{SF}_{\text{FLASH}}}{\text{SF}_{\text{conv}}} \sim \exp\left[\frac{\pi \eta_\Lambda n_c D_{O_2} p_c}{c_{\max}}\left(1 - \frac{c_{\max}}{c_{\max} + G_0(D/T)}\right)\right] \tag{SM5}$$

This is equivalent to Equation (12) in the main text with the identification $\eta \equiv \pi \eta_\Lambda$.

We note that the prediction in Equation (SM5) that the $\text{SF}_{\text{FLASH}}/\text{SF}_{\text{conv}}$ increases monotonically with increasing vascularity/oxygenation must eventually break down. Although this result is strictly valid for a random vasculature, it follows from the fact that (within this approximation) there exist avascular regions of arbitrarily large size, but with exponentially decreasing probability [Equation (SM1)]. Although we argue in the main text and in Supplementary Materials Section 4 that the vasculature exhibits pseudo-randomness over a range of inter-capillary distances, we hypothesize that the release of pro-angiogenesis growth factors and cell necrosis for anoxic cells ensure that there is a maximum distance-to-nearest-capillary value in actual tissues.

## 4. Comparison between simulated lattice and random capillary architectures with measured normal tissue distance to nearest capillary histograms

To evaluate the appropriateness of the random capillary array model, we calculated the distance-to-nearest-capillary (DNC) histogram for this geometry and compared it with the histogram calculated for a square lattice capillary array, as well as empirical results. The DNC histogram is a widely studied metric in animal normal and tumor tissue studies, the former including healthy and atrophied myocardium [10], skeletal muscle [11], brain [12], and skin [13], and is used to quantify the vascular architecture.

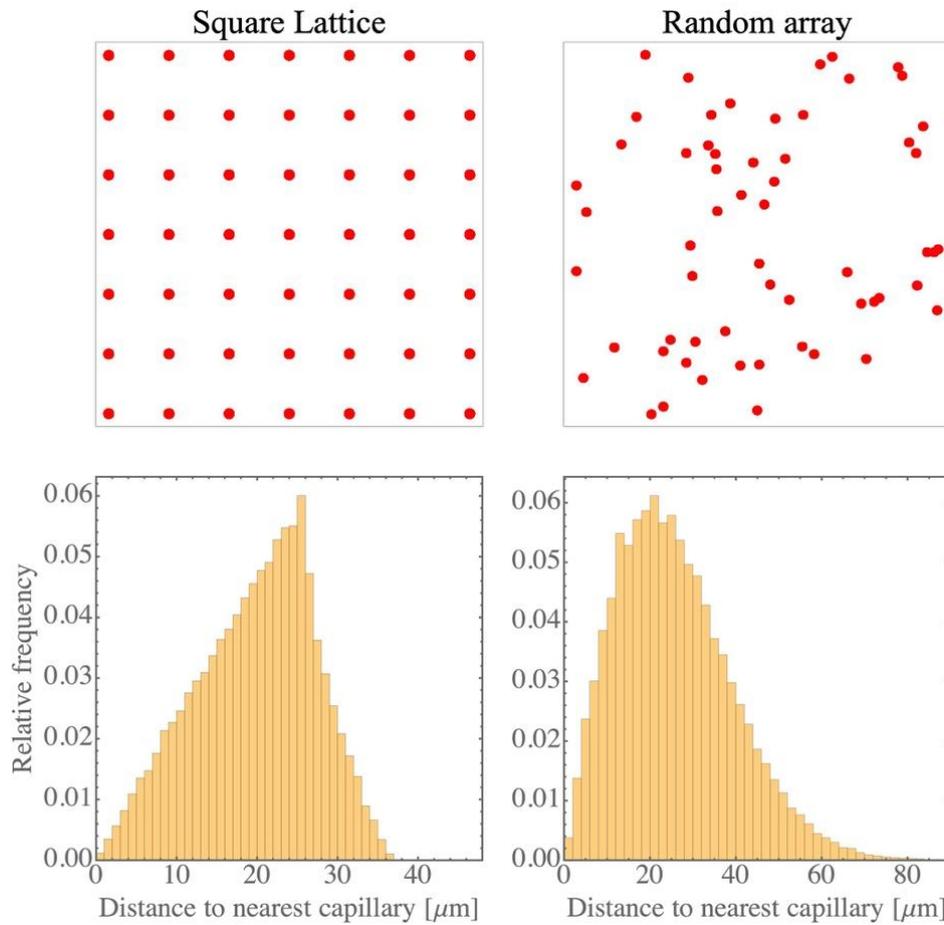

Figure SM2. Simulated distance to nearest capillary histograms for a square lattice (left) and random capillary array (right) for the same capillary density $n_c = 382$ mm$^{-2}$. The random capillary array histogram displays a long tail characteristic of observed normal and tumour tissue studies.

To calculate the DNC histograms, for a fixed capillary density $n_c$, two-dimensional domains were populated with capillaries, either randomly or at the lattice points of a square lattice (Figure SM2). Points in these domains were then randomly sampled; for each point, the distance to the nearest capillary was calculated and the resulting distribution of distances for the set of points was tabulated to give the DNC histogram. Figure SM2 (bottom panel) shows representative DNC histograms for square lattice and random capillary geometries, for $n_c = 382$ mm$^{-2}$.

A hallmark of empirically derived normal and tumor tissue histograms is a "long tail" in the distribution, extending well beyond the median DNC value [10, 12, 13]. Closely related to this feature is the long tail observed in histograms of the minimum intercapillary distance for healthy tissue [14-16]. This tail is evident in the DNC histogram calculated using the random capillary array panel (Figure SM2, right panel) but not in the histogram calculated using the lattice model (left panel), which displays an abrupt termination at the maximum DNC, $\text{DNC}_{\max} = 1/\sqrt{2n_c}$ (the distance from the corner to middle of each lattice unit cell). The existence of this tail, absent in the lattice capillary model, results in small regions of hypoxia, even in normal tissues [8]. Based on the similarities between the calculated and measured DNC histograms as well as the known presence of hypoxic regions in "well-perfused" normal tissues, we conclude that tissue vasculature architectures are better described by randomly allocated capillaries than a uniform capillary array.